\documentclass[prd,superscriptaddress,onecolumn,showpacs,11pt,msmath,preprintnumbers,showkeys]{revtex4}
\usepackage{wrapfig,rotating}
\usepackage{graphicx,epsfig,color}

\makeatletter
\newenvironment{tablehere}
{\def\@captype{table}}
{}
\newenvironment{figurehere}
{\def\@captype{figure}}
{}
\makeatother

\usepackage{graphicx}
\usepackage{dcolumn}
\usepackage{bm}

\def\beq{\begin{equation}}
\def\eeq{\end{equation}}
\def\beeq{\begin{eqnarray}}
\def\eeeq{\end{eqnarray}}

\def\2GPD{$_2\mbox{GPD}$}

\def\12{$1\otimes 2$}
\def\22{$2 \otimes 2$}

\def\Qsep{Q_{\mbox{\rm\scriptsize sep}}}
\def\Qsep2{Q^2_{\mbox{\rm\scriptsize sep}}}

\begin{document}
 \title{Heavy Quark Radiation in the Quark-Gluon Plasma in the Moliere Theory: Angular Distribution of the Radiation.}
 \author{B.\ Blok$^{1}$,\\[2mm] \normalsize $^1$ Department of Physics, Technion -- Israel Institute of Technology,
 Haifa, Israel\\}
 \begin{abstract}
\par We study the effects of adding the Coulomb interactions to the harmonic oscillator (HO) approximation of the heavy parton propagating through the quark-gluon plasma
(the extension to QCD of the Molliere theory).
We explicitly find  the expression for  the transverse momentum distribution of the gluon radiation of the heavy quark propagating in the quark gluon plasma in the framework of the Moliere theory,
taking into account the BDMPSZ radiation in the harmonic oscillator  (HO) approximation, and the Coulomb logarithms described by the additional logarithmic terms in the 
effective potential. We show that these  Coulomb logarithms significantly influence the  HO distribution, derived in the BDMPSZ works, especially for the  small transverse momenta, filling the dead cone,
and reducing the dead cone suppression of the  heavy quark radiation (dead cone effect). In addition we study the effect of the phase space constraints on the heavy quark energy loss,
and argue that taking into account  of both the phase space constraints and of the Coulomb gluons reduces the dependence of the heavy quark energy loss on its mas in the HO approximation.
 \end{abstract}
   \maketitle
 \thispagestyle{empty}

 \vfill

\section{Introduction.}
\par The energy loss of  a quark propagating in the quark-gluon plasma (QGP) was extensively studied in recent years in different approaches.
in  particular in the harmonic oscillator approximation, developed by   BDMPSZ  \cite{BDMPS1,BDMPS2,BDMPS3,BDMPS4,Z0A,Z0B,Z0C}, and in the GLV opacity expansion formalism \cite{GLV1,GLV2,GLV3}.
The harmonic oscillator approximation  (HO) enables taking control of the coherence effects in the QGP media, while the GLV expansion, although it includes N hard scatterings (the order N opacity expansion)
includes also the potentially large Coulomb effects.
\par The study of the heavy quark energy loss has been long recognized as an important phenomenological tool to diagnose the medium created 
in heavy ion collisions. Moreover, the heavy quark energy loss can be studied in different formalisms, like opacity expansion, harmonic oscillator approximation (mean field BDMPSZ), ADS/CFT. As a result the detailed 
understanding of heavy quark energy loss and diffusion is of a paramount importance for understanding the properties of quark-gluon plasma.
\par The heavy quark energy loss was first explicitly studied in \cite{DK} in the harmonic oscillator approximation , whose authors predicted significant decrease of the heavy quark energy loss and 
of the heavy quark quenching weights  due to the dead cone effect, similar to the dead cone effect in vacuum:
\beq
\omega \frac{dI^{\rm vac}}{d\omega dk_t^2}\sim \frac{\alpha_sC_F}{\pi^2}\frac{k_t^2}{(k_t^2+\theta^2\omega^2)^2},
\label{a1}
\eeq
where $m$ is the mass of the radiating  heavy quark, $E$ is the heavy quark energy, $\theta =m/E$ and  $k_t,  \omega$  are the transverse momenta and the energy of the radiated gluon.
\par  This effect  however was found to be in a disagreement with the experimental data
that shows   that quenching weights for heavy and light quark are very close up to rather small jet energies of 25-35 GeV\cite{wang,kurt}. This
contradiction led to an extensive research on  the heavy quark radiation in the quark gluon plasma.
\par One interesting question is whether the heavy quark energy loss mechanism can be studied in the pQCD framework. There were two approaches to this problem.
First the studies based on N=1 opacity expansion, second based on the mean field BDMPSZ mean field approach.
\par The studies based on the N=1 opacity expansion, starting from the pioneering works \cite{DG,ASW} eventually led to a number of realistic models models \cite{D1,D2,D3,D4,D5,D6,
Vitev1,Vitev2,Vitev3,Vitev4,Vitev5,Vitev6,majumder1,majumder2,majumder3,majumder4,ZHQ},
see also \cite{rev0,rev1,rev2,rev3} for recent reviews.
\par  The N=1 opacity expansion approach, although it accounts for some of the coherence and Coulombic interactions includes only single hard scattering \cite{Wiedemann3} . Thus we have another approach,
based on the extension of coherence effects connected with Landau Pomeranchuk Migdal (LPM) effect to the case of QCD media.
 This just the  the mean field BDSMPZ formalism.  The connection between BDMPSZ approach and the N=1 opacity expansion was studied in detail  in \cite{arnold1}. This approach
 was extended to heavy quark case in \cite{DK} and later heavy quark effects were studied in this framework in \cite{ASW,Z,BT} . In particular, in \cite{ASW} the important role of correct
 phase space restrictions was found, and in \cite{BT} the higher order corrections to heavy quark jet energy loss were studied.
\par  Note that up to recently only harmonic oscillator approximation for the BDMPSZ
 approach was developed. The inclusion of Coulombic Logarithms  in the BDMPSZ formalism (called the Moliere theory in the framework of conventional LPM effect in QED,
 (see i.e. \cite{katkov} for a review )
  was done only recently
in \cite{mehtar,mehtar1,mehtar2} for light quarks and gluons. The Moliere theory was extended to the case of heavy quarks in '\cite{blok1}.
The latter works however did not include  the study of the transverse momenta distributions, and the energy loss calculated without taking into account 
the proper phase constraints on the radiated gluons. Thus it is of great interest  to include the effects of Coulomb Logarithms in the transverse distributions in the 
BDMPSZ approach. This problem arises the special interest in the case of heavy quarks where the important role of phase constraints was first stressed in \cite{ASW}.
\par In this paper we extend the BDMPSZ  approach  for parton propagation in the QGP to include the Coulomb logarithms in the angular/transverse  momentum  distributions
for both massless and massive quarks. We shall explicitly calculate the form and the effects of Coulomb distributions in angular distributions of heavy (and light) quarks and estimate 
the combined influence of Coulomb effects and phase constraints on the heavy quark energy loss in the ASW framework. 
We shall see that Coulombic contribution is always positive and tends to increase the angular HO BDMPSZ distributions in the N=1 GLV direction.
\par Along the previous papers on the subject  \cite{mehtar,mehtar1,mehtar2} we shall make only  rather qualitative comparison to the experimental data,
and concentrate on model independent calculations.

\par The detailed 
comparison to the experimental data needs additional model dependent inputs like inclusion of the expansion, correct phase constraints (including realistic phase space constraints, taking into account 
 on the use of soft gluon approximation \cite{Wiedemann1},
general for BDMPSZ approach. Thus the detailed calculation of $v_2$ and $R_A$ for heavy quarks will be done elsewhere. will be done elsewhere.

\par The paper is organized in the following way. In chapter  2 we consider the basic formalism for calculation  the angular distribution of the radiation.
In chapter 3 we review the calculation in the BDMPS approach in the harmonic oscillator approximation \cite{Wiedemann1, ASW}, in section 4 we have the Moliere 
theory calculation, in section 5 we present numerical results for angular distributions, for corresponding energy loss and quenching weights in the soft gljuon approximation and integrating 
in transverse momenta $k_t\le \omega$, where $\omega$ is the energy of the radiated gluon..
In section 6 we take into account the 
energy conservation in the Leading Logarithmic Approximation, leading to improved phase space constraints for gluon radiation.
We see that the inclusion of these constraints leads to further decrease of the dependence of heavy quark energy loss
on its mass. Our results are summarised in conclusion.  Some useful mathematical formulae are given in the Appendix.
\section{Basic formalism}
\subsection{Basic formulae}
The heavy quark angular distribution in the media    is given by \cite{ASW}
 \begin{eqnarray}
 \omega \frac{dI}{d\omega d^2k }&=&\frac{C_F\alpha_s}{(2\pi)^2\omega^2}2Re\int d^2y \int ^\infty_0dt_1\int^{t_1}_0 dt e^{-i\vec k_t\vec y}\nonumber\\[10pt]
&\times& e^{-\int^\infty_{t_1} dsn(s)V(\vec y(s))}\partial_{\vec x}\partial _{\vec y}(K(\vec y,t_1,\vec x,t)-K_0(\vec y,t_1;\vec x,t))\vert_{\vec x=0}.\nonumber\\[10pt]
\label{e1}
 \end{eqnarray}
 Here $K$ is the propagator of the  particle in the media with the two dimensional effective potential due. to the scattering  centres, and 
 $K_0$ is the corresponding propagator of the free particle in the vacuum. The effective two dimensional potential is given by 
 \beq
 V(\vec\rho )=i\int \frac{d^2q_t}{(2\pi)^2}(1-\exp(i\vec q_t\vec\rho))\frac{d^2\sigma_{el}}{d^2q_t}.
 \eeq 
 Here $d^2\sigma_{el}/d^2q_t$ is the cross section of elastic scattering of high energy particle 
 on the media centre.   We assume the static media of the form 
 \beq
  n(s)=U(L-s)U (s)
  \eeq
  where $U=1$ if $s\ge 0$ and 0 if $s<0$ is a conventional step function.
 
 The  media is described by Gyulassy-Wang model  \cite{GW}.
 The effective potential in the momentum space is given by 
  \beq
\frac{ d\sigma(\vec q_t )}{d^2 q_t}=\frac{4\pi\alpha_sm^2_DT}{(q_t^2+\mu^2)^2}\equiv \frac{g^4n}{(q_t^2+\mu^2)^2},
 \eeq
 where  the parameter $\mu\sim m_D$, and the Debye mass $m_D$ is given by 
 \beq
 m_D\sim  4\pi\alpha_sT^2(1+N_f/6)=\frac{3}{2}g^2T^6
 \eeq
 for $N_f=3$ light quarks, T is the media/QGP temperature. The density of the scattering centres in the GW model is given by $n=\frac{3}{2}T^3$,
 and  the strong coupling is $\alpha_s=\frac{g^2}{4\pi}$.
 The effective potential in the coordinate space is 
  \beq
 V(\rho)=\frac{\hat q}{4N_c}(1-\mu \rho K_1(\mu \rho )=\frac{\hat q\rho^2}{4N_c}(\log(\frac{4}{\mu^2\rho^2})+1-2\gamma_E)\label{pot},
 \eeq 
 where $\gamma_E=0.577$ is the Euler constant,  and the bare quenching coefficient is
 \beq
 \hat q=4\pi\alpha_s^2N_cn.
 \eeq
 Note that $\hat q$ is fully determined by media properties, and does not depend on the quark mass. 
 \par  For processes that are dominated by large momentum transfer  is enough to
 take into account only the first terms in the Taylor expansion of $V(\rho)$. 
 The first approximation  corresponds to the quadratic term in the expansion \ref{pot} and is called the HO (harmonic oscillator ) approximation.
 In this approximation the effective potential V is given by
 \beq
 V(\rho)=\frac{1}{4}\hat q_{\rm eff}\rho^2\label{lpm}.
 \eeq
Here $\hat q_{\rm eff}$ is the  effective jet quenching coefficient, given by 
\beq
\hat q_{\rm eff}=\hat q\log( \frac{Q^2}{ \mu^2})\label{eff},
\eeq
and Q is the typical transverse momenta, accumulated by the particle on the scale of the coherence length.

\par The HO effectively describes the LPM bremsstrahlung \cite{BDMPS1}. More 
precise treatment of the energy loss includes also large Coulomb  logarithms and is called in the theory 
of the Abelian (QED) LPM effect the Moliere theory \cite{katkov}. In the QCD framework the inclusion of Coulomb interactions 
can be made using the perturbation theory \cite{mehtar,mehtar1}. Namely, instead of the usual opacity expansion
\cite{GLV1,GLV2,GLV3}, we shall consider the perturbation theory around the oscillator potential adding the Coulomb effects as a perturbation.
The effective potential in Moliere theory is given by 
\beq
V(\rho) =\frac{1}{4}\hat q\rho^2\log(1/\rho^2\mu^2),
\eeq
and includes the short range Coulomb  logarithms. In the framework of the perturbation theory this potential is split as 
\beq
V(\rho)=V_{HO}(\rho)+V_{pert}(\rho), V_{HO}(\rho)=\frac{\hat q\log(Q^2/\mu^2)}{4}\rho^2,V_{\rm pert}(\rho)=\frac{\hat q}{4}\log(\frac{1}{Q^2\rho^2}),\label{pert}
\eeq
where Q is the typical momenta, defined above, equal to $Q\sim \sqrt{\hat q\omega}$ in the HO approximation.
We shall need sufficiently large Q, so that 
\beq
\log(Q^2/\mu^2)\gg \log(\frac{1}{Q^2\rho^2}),
\eeq
i.e. perturbation theory is applicable meaning that we probe rather small transverse distances.
\par Then the energy loss is given by  Eq. \ref{e1},
where the propagator K is calculated in perturbation theory as  \cite{mehtar,mehtar1}
\beq
K(\vec y,t_1;\vec x,t)= K_{HO}(\vec y,t_1;\vec x,t)-\int d^2z\int^{t_1}_t dsK_{HO}(\vec y,t_1;\vec z,s) V_{pert}(z)K_{HO}(\vec z,s;\vec x,t)\label{15}
\eeq
Here $K_{HO}$ is the heavy  quark propagator in the imaginary two dimensional potential $V_{HO}$    \cite{ASW}:

\begin{eqnarray}
K_{HO}(\vec y,t_1;\vec x,t)&=&\frac{i\omega\Omega}{2\pi \sinh \Omega (t_1-t)}\exp(\frac{i\omega 
\Omega }{2}\{\coth\Omega  (t_1-t)(\vec x^2+\vec y^2)-\nonumber\\[10pt]
&-&\frac{2\vec x\vec y }{\sinh\Omega (t_1-t)}\})\exp(-i\theta^2\omega (t_1-t)/2),\nonumber\\[10pt]
\label{e2}
\end{eqnarray}
and 
\beq
\Omega=\frac{(1+i)}{2}\sqrt{\frac{\hat q}{\omega}}
\eeq
In the limit  when there is no media this propagator reduces to free quark propagator 
\beq
K_0(\vec y,t_1;\vec x,t)=\frac{i\omega}{2\pi }\exp(i\frac{\omega (\vec x-\vec y)^2}{2(t_1-t)}).\label{e3}
\eeq
\subsection{Qualitative Dynamics of the Heavy Quark}
\par The expansion written in the form \ref{15} clearly exhibits the formation lengths described in the Introduction:
the heavy quark mass leads to the oscillating exponent $\exp(i\theta^2\omega/2 (t_1-t))$ in Eq. \ref{e2}, while the harmonic oscillator part of the propagator \ref{e2} oscillates 
with the frequency $\sqrt{\omega/\hat q}$. Then it  is clear that when $l_c^q<< l_c^{LPM}$ the oscillations due to heavy quark mass cut off the integral for heavy quark
energy loss, the  oscillating harmonic oscillator part of the propagator is approximately freezed and the LPM effect is not relevant, the energy
loss is defined by the induced radiation on the scattering centres-the N=1 GLV. On the other hand, in the opposite case, the heavy quark exponent
is close to one, and the integral for energy loss is controlled by the HO multiplier. We have LPM bremsstrahlung plus corrections due to  Coulomb logarithms.
\par We can now choose the subtraction scale $Q$ in the momentum space. As it was explained in \cite{arnold1,mehtar}  this scale corresponds to the typical momentum accumulated 
by the quark along the coherence length propagation. Such momentum squared  is $\hat q\times\sqrt{\omega/\hat q}$ for $\omega<<\omega_{DC}$ and $\sim \theta^2\omega^2\sim \omega/l^q_c$
for $\omega>>\omega_{DC}$. Consequently we shall use the interpolation formula
\beq 
Q^2=\sqrt{\omega \hat q_{\rm eff}}U(-\omega+\omega_{DC})+\theta^2\omega^2U(\omega -\omega_{DC}),\label{m1}
\eeq
where $U(x)$ is a unit step function:$U(x)=1$ if $x\ge 0$, and $U(x)=0$ if $x\le 0$.
\par Alternatively, the dynamics of the heavy quark can be approached using the arguments in \cite{arnold1}. Namely , in the LPM (diffusion ) regime the distribution
over momentum transfers in the scattering on the media centres is described by a gaussian, peaked in the $Q^2_{typ}\sim \sqrt{\hat q w}$. The scattering with significantly higher 
momentum transfers $q_t$ is described by the tail of the distribution, which is N=1 GLV, that essentially describes the independent  scattering on the media centres.
In this region the LPM gaussian is parametrically close to zero, and N=1 GLV dominates. It was explained in \cite{Z1} that N=1 term in opacity expansion is a good description of large momentun transfer regime, since such
scatterings in the tail occur quite rarely. Since inside dead cone the typical momenta is  $k^2_t\sim \omega/l_c^q\sim \theta^2\omega^2\gg \sqrt{\hat q\omega}$, inside the dead cone we shall find ourselves in 
the GLV regime.
\subsection{N=1 GLV}
\par We shall also need the explicit expression for N=1 term in the opacity expansion for  angular distribution for massive quark. The corresponding result was 
derived in \cite{ASW}, and has the form:
\begin{eqnarray}
\omega \frac{dI}{d\omega d^2k_t} &=&\int_0^\infty dq^2\frac{2\alpha_sC_F\hat q }{\pi^2\omega}\frac{ LQ_1-\sin(LQ_1)}{Q_1^2}\frac{q^2}{q^2+\theta^2\omega^2}\nonumber\\[10pt]
&\times&\frac{m^2_D(k^2+\theta^2\omega^2)+(k^2-\theta^2\omega^2)(k^2-q^2)}{(k^2+\theta^2\omega^2)((m^2+k^2+q^2)^2-4k^2q^2)^{3/2}}.\nonumber\\[10pt]
\label{GLV}
\end{eqnarray}
where 
\beq
Q_1=(q^2+\theta^2\omega^2)/(2\omega).
\eeq
Here $k_t$ is the momentum of the radiated gluon.
\section{Angular distribution in the Harmonic Oscillator Approximation.}
\par The angular distribution of the gluon radiation was first calculated for heavy quark in \cite{ASW} , and contains two contributions: The first is the bulk contribution and is given by 
 \begin{eqnarray}
 \omega \frac{dI^{HO\,\,\,Bulk}}{d\omega d^2k_t }&=&\frac{\alpha_sC_F}{(2\pi)^2\omega^2}2Re\int d^2y \int ^L_0dt_1\int^{t_1}_0 dt e^{-i\vec k_t\vec y}\nonumber\\[10pt]
&\times& e^{-1/4\hat q (L-t_1)y^2}\partial_{\vec x}\partial _{\vec y}(K(\vec y,t_1,\vec x,t)-K_0(\vec y,t_1;\vec x,t))\vert_{\vec x=0}.,\nonumber\\[10pt]
\label{e1a}
 \end{eqnarray}
where K is the heavy quark propagator in harmonic oscillator approximation given by Eq. \ref{e2}.
\par The second contribution is a boundary term given by 
 \begin{eqnarray}
 \omega \frac{dI^{\rm HO\,\,\, boundary}}{d\omega d^2k _t}&=&\frac{\alpha_sC_F}{(2\pi)^2\omega^2}2Re\int d^2y \int ^\infty_Ldt_1\int^{L}_0 dt e^{-i\vec k_t\vec y}\nonumber\\[10pt]
&\times& \partial_{\vec x}\partial _{\vec y}(K(\vec y,t_1,\vec x,t)-K_0(\vec y,t_1;\vec x,t))\vert_{\vec x=0},\nonumber\\[10pt]
\label{e1b}
 \end{eqnarray}

where the propagator K is given by  ($t_1>L>t$)
\beq
K(\vec y,t_1;\vec x ,t)=\int d^2z K_0(\vec y,t_1;\vec z,L)K_{HO}(\vec z,L;\vec x ,t).
\eeq
The direct calculation shows that the bulk  term is given by:
\begin{eqnarray}
 \omega \frac{dI^{\rm HO\,\,\, Bulk}}{d\omega d^2k_t }&=&-2{\rm Re}\int^L_0dt\int^L_tdt_1 \frac{\alpha_sC_F\Omega^2}{\pi^2R^2\sinh{\Omega(t_1-t)}^2}(q(L-t_1)-\frac{2i\omega\Omega\cosh{\Omega(t_1-t)}}{R})\nonumber\\[10pt]
&\times& \exp(i\theta^2\omega(t-t_1)/2)\exp(-k_t^2/R),\nonumber\\[10pt]
\label{k1}
 \end{eqnarray}
 where 
 \beq
 R=q(L-t_1)-2i\omega\Omega\coth{\Omega (t_1-t)},
 \eeq
 The boundary term is given by 
 \beq
  \omega \frac{dI^{\rm HO\,\,\, boundary}}{d\omega d^2k _t}=\int ^L_0dt\frac{-i\alpha_sC_Fk^2_t}{(k_t^2+\theta^2\omega^2)(2\pi)^2\omega}\frac{\exp(\frac{-ik^2_t\tanh\Omega(L-t)}{2\omega\Omega})\exp(i\theta^2\omega(t-L)/2}{\cosh \Omega(L-t)^2}
  \label{g1}
  \eeq
 from these expressions we subtract their $\hat q=0$ limit.
 These expressions of course coincide with the corresponding ones  in \cite{ASW}.
 \section{Coulomb corrections.}
Let us consider now the full expression for angular distribution:
 \begin{eqnarray}
 \omega \frac{dI}{d\omega d^2k_t }&=&\frac{\alpha_sC_F}{(2\pi)^2\omega^2}2Re\int d^2y \int ^\infty_0dt_1\int^{t_1}_0 dt e^{-i\vec k_t\vec y}\nonumber\\[10pt]
&\times& e^{-\int^\infty_{t_1} dsn(s)(V_{HO}+V_{pert})(\vec y(s))}\partial_{\vec x}\partial _{\vec y}(K(\vec y,t_1,\vec x,t)-K_0(\vec y,t_1;\vec x,t))\vert_{\vec x=0}.\nonumber\\[10pt]
\label{e1d}
 \end{eqnarray}
where $K $ is now the full propagator that is also calculated in the perturbation  theory:
\beq
K=K_{HO}+K_{HO}V_{pert}K_{HO}
\eeq
 In the Moliere theory approach we carry the perturbation theory over $V_{pert}$ with the solution for harmonic oscillator approximation being the zero order term. Then it is clear from Eq. \ref{e1d} that there are two 
 distinct term in the perturbation theory: first term is due to the expansion of the exponent in Eq. \ref{e1d} in powers of $V_{pert}$, while the second term is due to expansion of the propagator. The latter term is in turn a sum of two terms,
 first the boundary term with $t_1>L$ and the bulk term with $t_1<L$. We shall now move to calculation of these 3 terms: the term that comes from the exponent expansion and the two terms that come from the perturbative expansion of the propagator.
 \subsection{Exponent expansion}
 \par  Explicitly this term is given by 
 \begin{eqnarray}
   \omega \frac{dI^{\rm Coulomb\,\,\, one}}{d\omega d^2k }&=&-\frac{\alpha_sC_F}{(2\pi)^2\omega^2}2Re\int d^2y \int ^L_0dt_1\int^{t_1}_0 dt e^{-i\vec k_t\vec y}\nonumber\\[10pt]
&\times& (L-t_1)V_{pert}(\vec y)\partial_{\vec x}\partial _{\vec y}(K_{HO}(\vec y,t_1,\vec x,t)-K_0(\vec y,t_1;\vec x,t))\vert_{\vec x=0}.\nonumber\\[10pt]
\label{e1n}
   \end{eqnarray}
  Substituting the known expressions for  $V_{pert}$ and the propagators we obtain
  \begin{eqnarray}
   \omega \frac{dI^{\rm Coulomb\,\,\, one}}{d\omega d^2k }&=&-\frac{\hat q}{4}\alpha_sC_F\int^L_0dt_1\int ^{t_1}_0dt \frac{\Omega^2}{(2\pi)^3\sinh(\Omega (t_1-t)^2}\nonumber\\[10pt]
   &\int &d^2u u^2\log(1/(u^2Q^2))(2+i\omega\Omega\coth (\Omega(t_1-t) ) u^2)\exp(i\omega \Omega\coth(\Omega (t_1-t))u^2/2)\nonumber\\[10pt]
   &\times&\exp(-\frac{1}{4} \hat q (L-t_1)u^2-i\vec k_t\vec u)\exp(i\theta^2\omega  (t-t_1)/2).\nonumber\\[10pt]
   \label{c1}
   \end{eqnarray}
   The integral over the transverse momenta azimutal angle can be easily taken using the standard integral  \cite{AS}
   \beq
   \int^{2\pi}_0d\phi \exp(-i\vec k_t\vec u)=J_0(k_tu)
   \eeq
   Let us introduce two new functions that can be expressed through elementary functions (see appendix A):
   \begin{eqnarray}
   F_2(p,c,Q)&=&\int^\infty_0x^3\log(x^2Q^2)J_0(cx)\exp(-px^2)\nonumber\\[10pt]
   F_3(p,c,Q)&=&\int^\infty_0x^5\log(x^2Q^2)J_0(cx)\exp(-px^2).\nonumber\\[10pt]
   \end{eqnarray}
   So we finally get 
   \begin{eqnarray}
    \omega \frac{dI^{\rm Coulomb\,\,\, one}}{d\omega d^2k_t }&=&\frac{\hat q}{4}2Re\frac{\alpha_sC_F\Omega^2}{(2\pi)^2}\int^L_0 dt_1\int_0^{t_1}dt (2F_2(R/4,k_t,Q)+i\omega\Omega\coth(\Omega (t_1-t))F_3(R/4,k_t,Q))\nonumber\\[10pt]
    R&=&=\hat q (L-t_1)-2i\omega\Omega \coth\Omega(t_1-t)\nonumber\\[10pt]
    \label{R}
    \end{eqnarray}
   \subsection{Propagator expansion:the bulk term}
   We now consider the contribution to the angular distribution due to the perturbative expansion of the propagator in the powers of $V_{pert}.$.
   We have in the integral over $t_1$ two terms: the first is from 0 to L and is called a bulk term, the second corresponds to the case  when $t_1>L$ and is called a boundary term. Let us consider first the bulk term
   \begin{eqnarray}
   \omega \frac{dI^{\rm Coulomb\,\,\, bulk}}{d\omega d^2k }&=&\frac{\alpha_sC_F}{(2\pi)^2\omega^2}\int d^2z\int d^2u\int^L_0 dt_1\int ^{t_1}_0dt\int^{t_1}_t ds \frac{e^{-\hat q (L-t_1)u^2/4-i\vec  k_t\vec u}}{\sinh \Omega(t_1-s)\sinh\Omega (s-t)}\nonumber\\[10pt]
   &\times&\partial_{\vec y}\partial_{\vec u}K_{HO}(u,t_1;\vec z,s)(\frac{\hat q}{4}z^2\log(1/z^2Q^2))K_{HO}(\vec z,s;t,\vec y =0)\nonumber\\[10pt]
   \end{eqnarray}
   After integration by parts we obtain, calculating the gaussian integral over $d^2u$
  \begin{eqnarray}
     \omega \frac{dI^{\rm Coulomb\,\,\,  bulk}}{d\omega d^2k _t}&=&\int d^2z\int^L_0dt_1\int^{t_1}_0 dt\int^{t_1}_t ds
     \frac{-\alpha_sC_Fi\omega^2\Omega^4(\hat q (L-t_1)z^2+2i\cosh (\Omega (t_1-s)\vec k_t\vec z)}{16\pi^3\sinh(\Omega(t_1-s))^2\sinh(\Omega (s-t))^2}\nonumber\\[10pt]
     &\times& \frac{z^2\log(1/(z^2Q^2))}{R(t_1,s)^2}\exp(-k^2_t/R(t_1,s)-2\omega\Omega\vec k_t\vec z/(R(t_1,s)\sinh\Omega (t_1-s)))\nonumber\\[10pt]
     &\times&\exp(i\frac{\omega\Omega z^2\sinh\Omega(t_1-t)}{2\sinh (\Omega (t_1-s))\sinh(\Omega (s-t))}\frac{R(t_1,t)}{R(t_1,s)})\nonumber\\[10pt]
     \label{b2}
\end{eqnarray}   
where $R(t_1,t)$ is given by Eq. \ref{R}.
The angular integral can be easily taken using the standard formulae \cite{AS}:
\begin{eqnarray}
\int^{2\pi}_0\cos(x)\exp(-A\cos(x))dx&=&2\pi iJ_1(iA)\nonumber\\[10pt]
 \int^{2\pi}_0\exp(-A\cos(x))dx&=&2\pi J_0(iA)\nonumber\\[10pt]
 \label{b3}
\end{eqnarray}
where $J_0,J_1$ are the conv entional Bessel functions.
Introducing an additional function
\beq
F_4(p,c,Q)=\int^\infty_0dz  z^4J_1(cz)\exp(-pz^2)\log(z^2Q^2)
\label{d4}
\eeq
we obtain final answer:
  \begin{eqnarray}
     \omega \frac{dI^{\rm Coulomb\,\,\, bulk}}{d\omega d^2k_t }&=&\alpha_sC_F\int^L_0dt_1\int^{t_1}_0dt\int^{t_1}_t ds\frac{q}{4}\frac{i\omega^2\Omega^4}{2\pi^2}\frac{\exp(-k_t^2/R(t_1,s))}{\sinh\Omega (t_1-s)^2\sinh \Omega (s-t)^2}\nonumber\\[10pt]
     &\times&(\hat q (L-t_1)F_3(p,c,Q)-2k_t\cosh \Omega (t_1-s)F_4(p,c,Q))\exp(i\theta^2\omega (t-t_1)/2)\nonumber\\[10pt]
     p&=&-\frac{\omega\Omega\sinh \Omega (t_1-t)}{2\sinh \Omega (t_1-s)\sinh \Omega (s-t)}\frac{R(t_1,t)}{R(t_1,s)}\nonumber\\[10pt]
     c&=&k_t\frac{2i\omega\Omega}{R(t_1,s)\sinh \Omega (t_1-s)}\nonumber\\[10pt]  
     \label{b4}   
     \end{eqnarray}
   \subsection{Boundary term.}
   
   Finally we consider the boundary contribution:
    \begin{eqnarray}
   \omega \frac{dI^{\rm Coulomb\,\,\, boundary}}{d\omega d^2k_t}&=&\frac{\alpha_sC_F}{(2\pi)^2\omega^2}\int d^2z\int d^2r\int d^2u\int^\infty_L dt_1\int ^{L}_0dt\int^{L}_t ds \frac{\exp(-i\vec  k_t\vec u )}{\sinh \Omega(t_1-s)\sinh\Omega (s-t)}\nonumber\\[10pt]
   &\times&\partial_{\vec y}\partial_{\vec u}K_0(t_1,\vec u;,L,\vec r)K_{HO}(\vec r,L;\vec z,s)(\frac{\hat q}{4}z^2\log(1/z^2Q^2)K_{HO}(\vec z,s;t,\vec y =0)\nonumber\\[10pt]
   \end{eqnarray}
We first do integral over $d^2y$ and over $t_1-L$,and then gaussian integral over $d^2r$. Using the Eqs. \ref{b3} for angular integration and the definition \ref{d4} of the function $F_4$
we easily obtain a final answer for the boundary term.
    \begin{eqnarray}
   \omega \frac{dI^{\rm Coulomb\,\,\, boundary}}{d\omega d^2k_t }&=&\frac{\hat q}{4}2\alpha_sC_FRe\frac{\omega \Omega^2}{(2\pi)^2}\frac{ik_t}{k_t^2+\theta^2\omega^2}\exp(-i\frac{k_t^2\tanh(\Omega (L-s))}{2\omega\Omega})\nonumber\\[10pt]
   &\times&F_4(\frac{-i\omega\Omega\cosh(\Omega (L-t)}{\cosh(\Omega (L-s)\sinh \Omega (s-t))}, \frac{k_t}{\cosh \Omega (L-s)},Q)\exp(i\theta^2\omega (t-L)/2). \nonumber\\[10pt]
   \label{b10}
\end{eqnarray}   
Note that functions $F_2,F_3,F_4$ can be easily expressed through known special functions,
the explicit expressions are given in the Appendix. In addition it is easy to perform integral over the transverse momentum $k_t$ between 0 and some scale $\omega_1$ analytically.

\section{Numerics}
\par Our final answer is the sum of all  terms that we calculated in the previous two chapters.
\beq
 \omega \frac{dI(\omega,\hat q,\theta,\mu ,L)}{d\omega d^2k_t }= \omega \frac{dI^{HO}(w,q_{eff},\theta,\mu ,Q_{eff} ,L)}{d\omega d^2k_t }+\frac{\hat q}{4} \omega \frac{dI^{\rm Coulomb}(\omega q_{eff},\theta,Q_{eff},\mu ,L)}{d\omega d^2k_t }\label{pn1}
 \eeq
 \beq
  \omega \frac{dI^{HO}}{d\omega d^2k_t }= \omega \frac{dI^{\rm HO\,\,\, Bulk}}{d\omega d^2k _t}+ \omega \frac{dI^{\rm HO\,\,\, Boundary}}{d\omega d^2k_t }
  \label{pn2}
  \eeq
  \beq
  \omega \frac{dI^{\rm Coulomb}}{d\omega d^2k_t }=\omega \frac{dI^{\rm Coulomb\,\,\, one}}{d\omega d^2k_t }+\omega \frac{dI^{\rm Coulomb\,\,\, Bulk}}{d\omega d^2k _t}+\omega \frac{dI^{\rm Coulomb\,\,\, Boundary}}{d\omega d^2k_t }
\label{pn3}
\eeq

where the terms with the index HO are given by Eqs.  \ref{k1}, \ref{g1}, while the terms with the index   are given by Eqs. \ref{b4},\ref{b10},\ref{R} (without external multiplier $\hat q$/4).
The effective scale $Q_{eff}$ is given by Eq. \ref{m1}, and the effective quenching coefficient is given by Eq. \ref{eff}.
\par For our numerical estimates we shall use the same parameters for QGP as in \cite{mehtar,mehtar1,blok1} :T =0.4 GeV, $\alpha_s =0.3$, leading to $\mu=m_D=0.9$ GeV and $\hat q =0.3$ GeV$^3$.
\par We do numerically double and triple integrals in $t,t_1,s$ using the Mathematica 12 software.

\subsection{Angular distributions for soft gluons.}

\par  We shall now present the numerical estimates for the  angular distributions of the radiated gluons and compare them with the BDMPS angular spectrum \cite{ASW} and N=1 GLV angular distributions.
We shall depict these angular distributions for typical values of 
:$\omega=5,10,20$ GeV in Figs. \ref{f1},\ref{f2},\ref{f3}. The BDMPS maximum angle is $\theta_{BDMPS}=(\hat q/\omega^3)^{1/4}=0.22$ for $\omega=5$ GeV,
0.13 for $\omega=10 $ GeV, and 0.08 for 20 GeV. for $\hat q\sim 0.3$ that we use in our calculations.
\par For  $\omega=5 $ GeV the BDMPS angle is outside the dead cone for all values of dead cone angle that we we consider,i.e. $\theta\le 0.2$. The Coulomb corrections to BDMPS are significant
and  are the biggest for small $k_t$, although the calculations for large $k_t$ especially of order $\omega$ are not trustworthy, since we use  soft gluon  approximation
in the BDMPS approach.  We see that the Coulomb correction is approximately constant at small $k_t$ and starts to decrease in parallel with BDMPS contribution.

\vskip 0.8 cm
\begin{figurehere}
\includegraphics[scale=0.5]{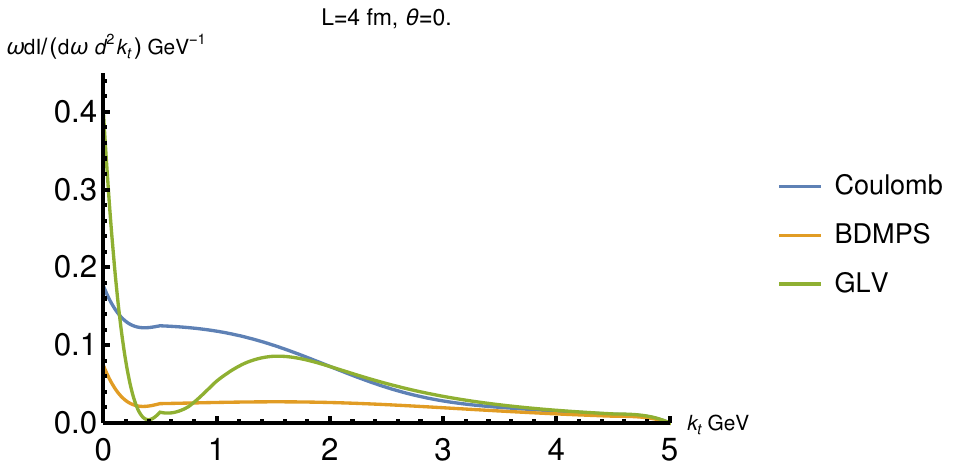}
\includegraphics[scale=0.5]{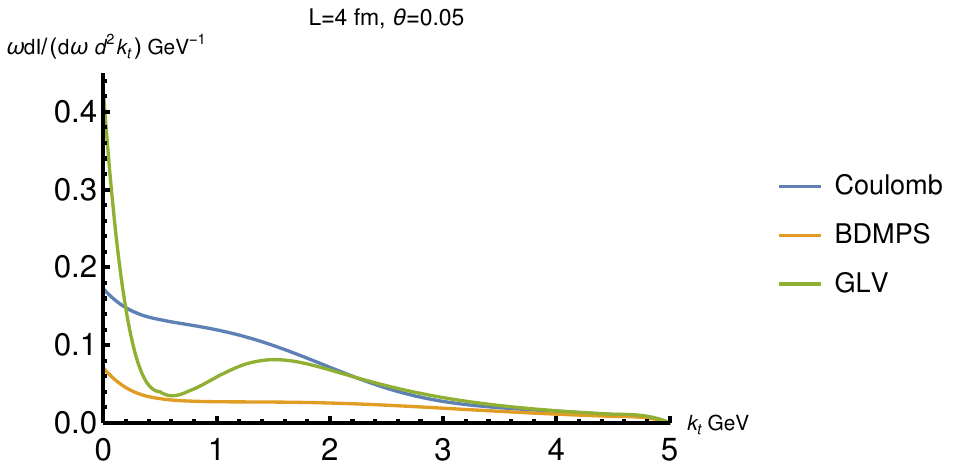}
\vskip 0.8 cm
\includegraphics[scale=0.5]{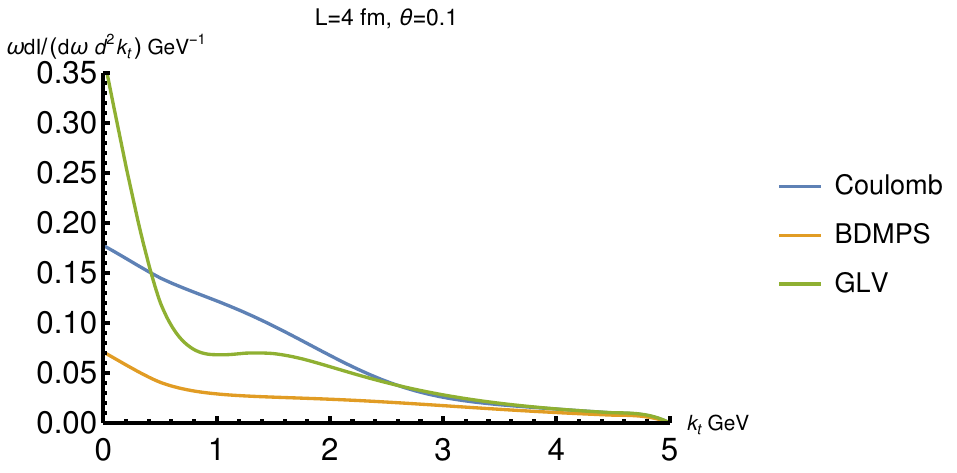}
\includegraphics[scale=0.5]{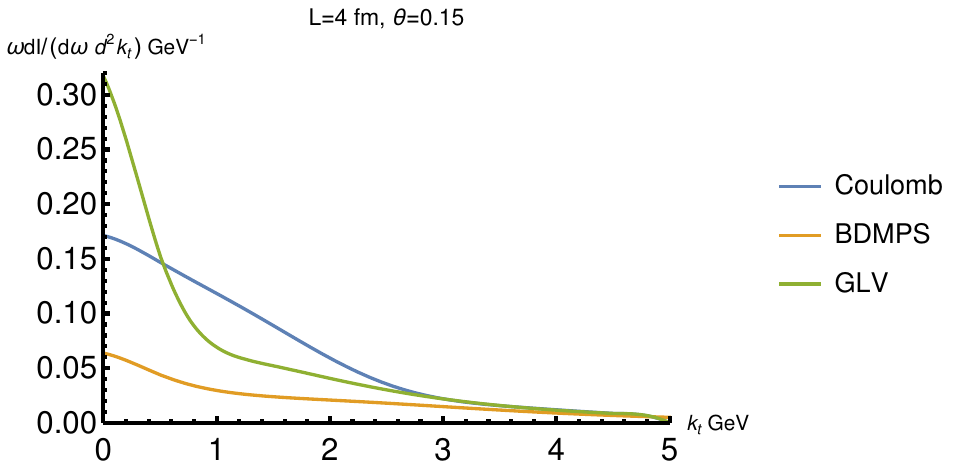}
\vskip 0.8 cm
\center{\includegraphics[scale=0.53]{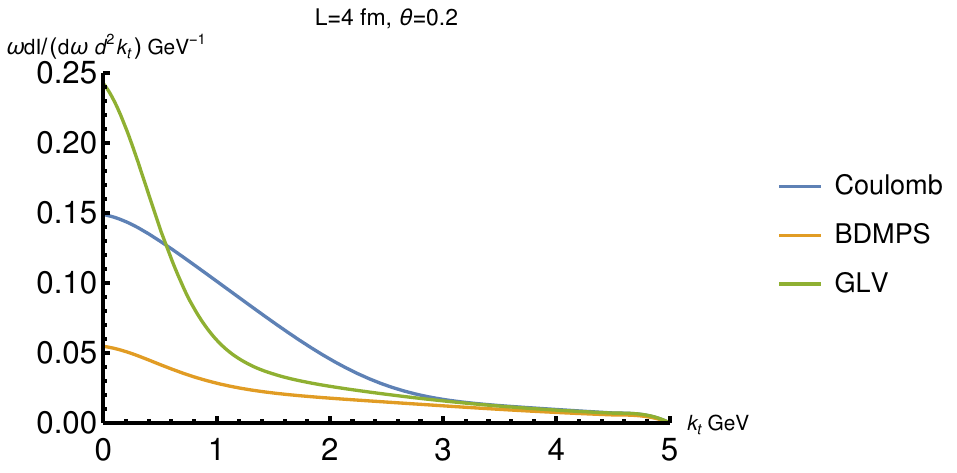}}
\caption{\textsl{The angular distribution of radiated gluons for $\omega= 5 $ GeV } for different $\theta=0,0.05,0.1,0.15,0.2$,  Here and in the  Figs. \ref{f1},\ref{f2} BDMPS means the BDMPS angular distribution
in the Harmonic Oscillator approximation  given by Eq. \ref{pn2},  Coulomb means the angular distribution in the Moliere theory  given by 
 Eq. \ref{pn1}, All graphs here and below are presented divided by $\alpha_sC_F$}. 
\label{f1}
\end{figurehere}
\vskip 1cm
\begin{figurehere}
\includegraphics[scale=0.5]{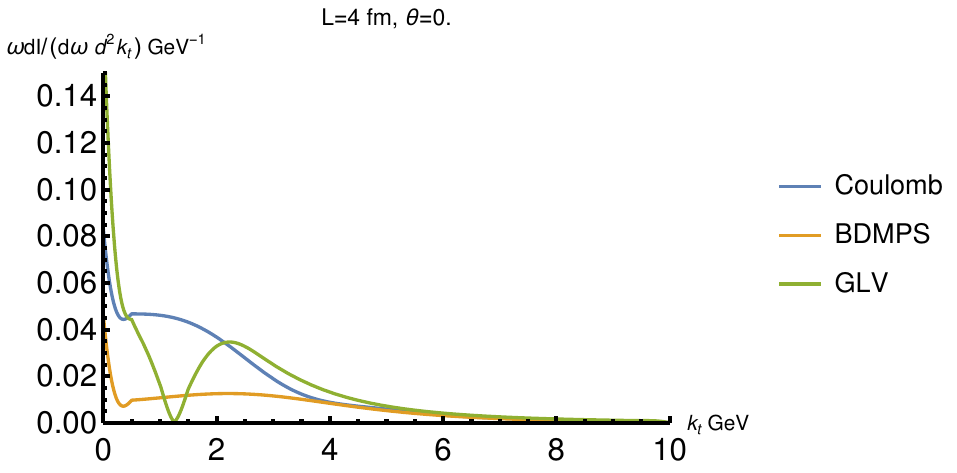}
\includegraphics[scale=0.5]{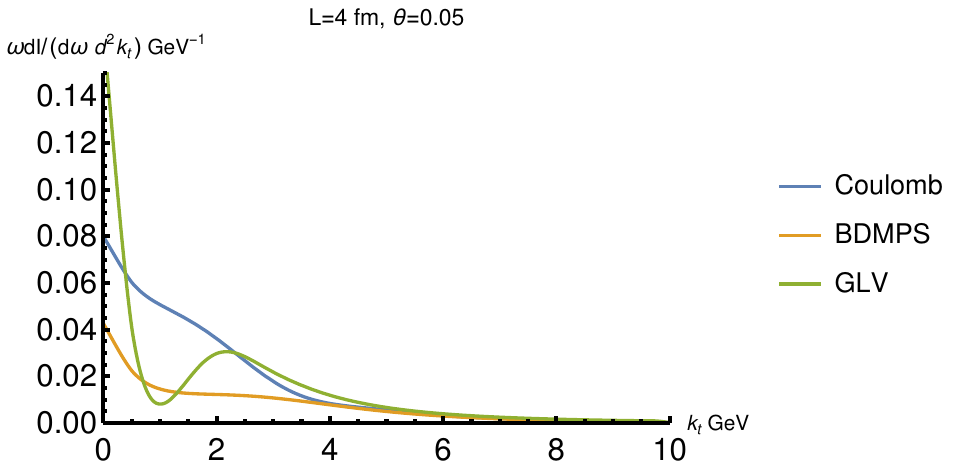}
\vskip 0.8 cm
\includegraphics[scale=0.5]{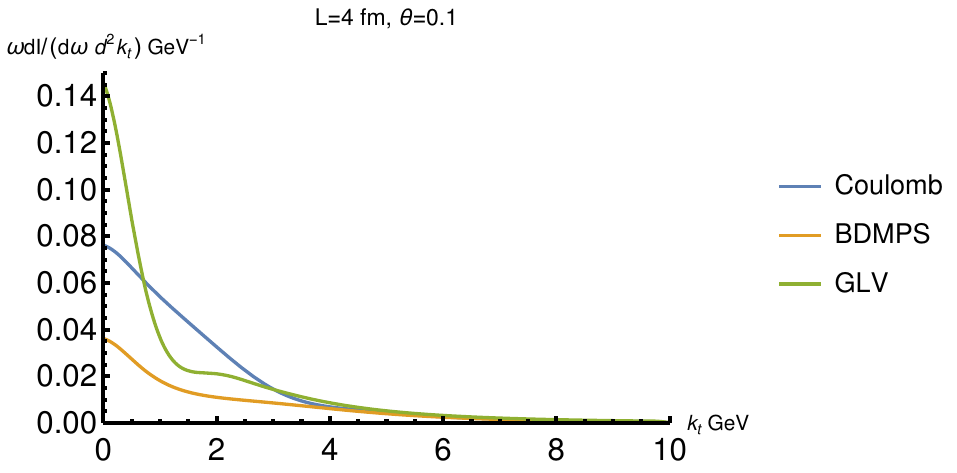}
\includegraphics[scale=0.5]{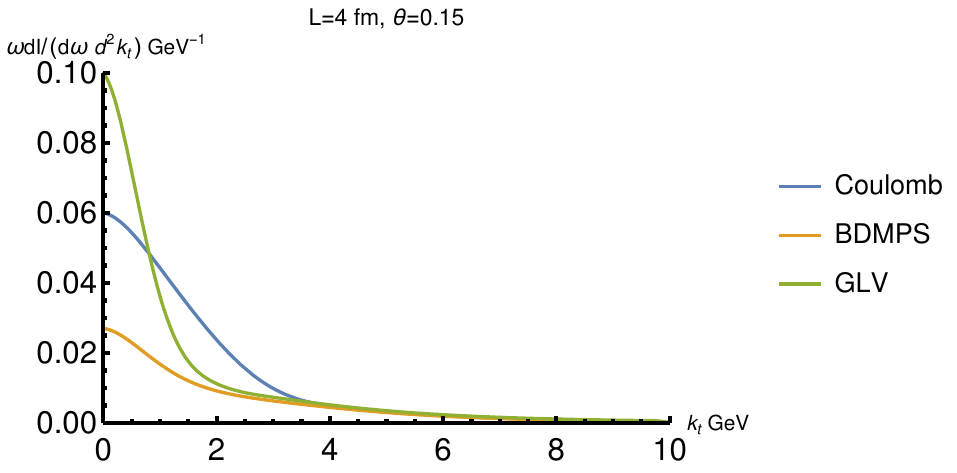}
\vskip 0.8 cm
\center{\includegraphics[scale=0.53]{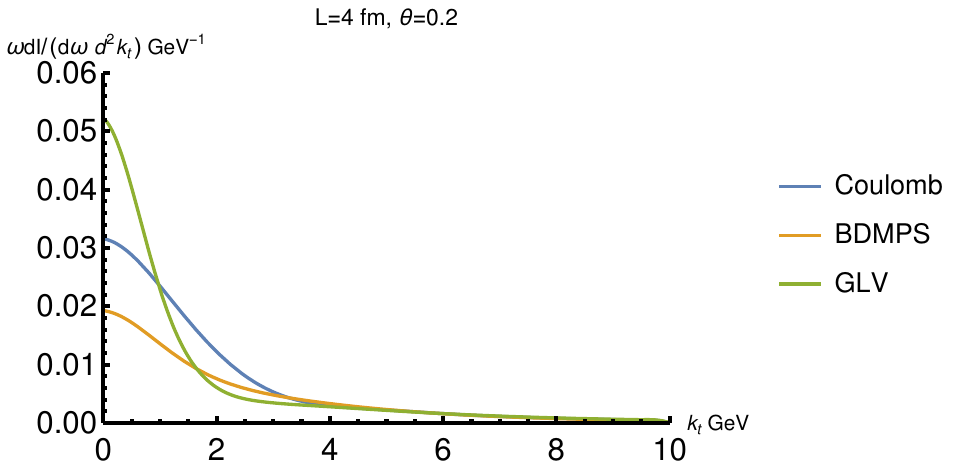}}
\caption{\textsl{The angular distribution of the radiated gluons for $\omega= 10 $ GeV } for different $\theta=0,0.05,0.1,0.15,0.2$}
\label{f2}
\end{figurehere}
\par In Fig. \ref{f2} we consider $\omega=10 $GeV. In this case two last values of  $\theta=0.15,0.2$ correspond to the BDMPS maximum. $\theta_{BDMPS}$  inside the dead cone. We see that.  the
Coulomb correction together with BDMPS gluons 
fill the dead cone.
\par In Fig.  \ref{f3}  already three last values of  $\theta=0.1,0.15,0.2$  correspond to the situation when $\theta_{BDMPS}$ is inside the dead cone. In all these cases there is no dead cone effect, and BDMPS and Coulomb radiation fills the dead cone region.

\vskip 1cm
\begin{figurehere}
\includegraphics[scale=0.5]{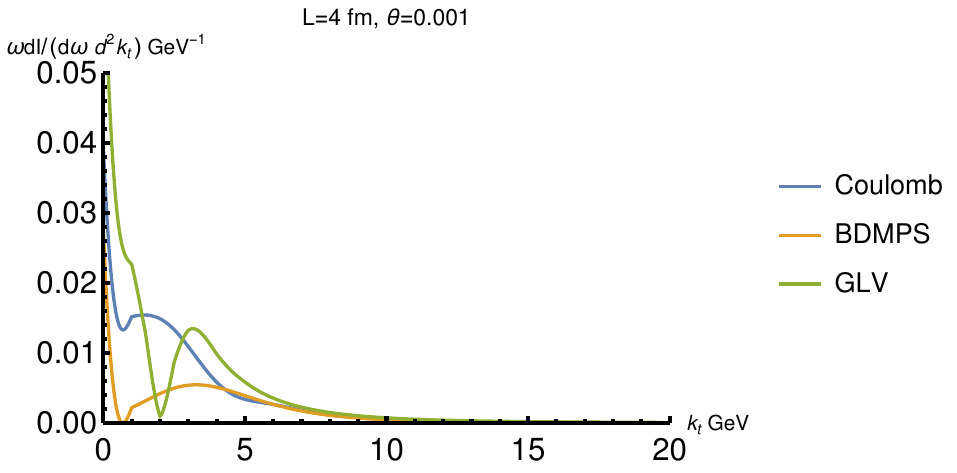}
\includegraphics[scale=0.5]{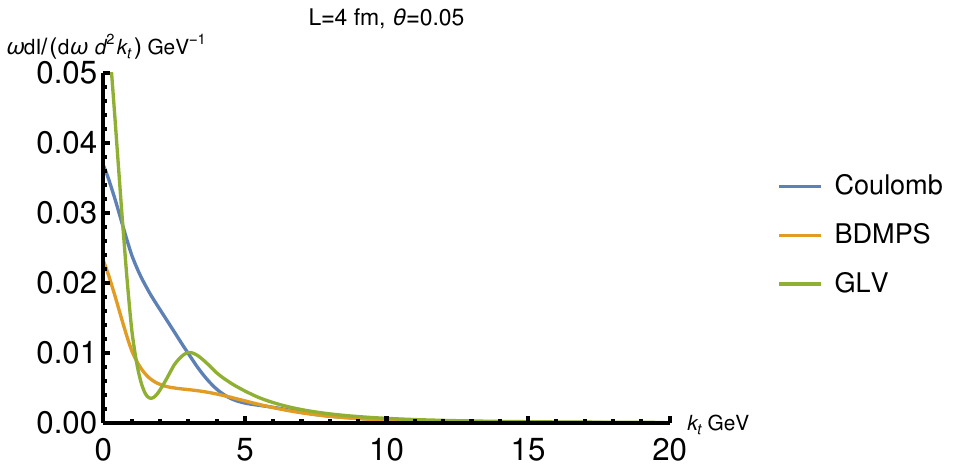}
\vskip 0.8 cm
\includegraphics[scale=0.5]{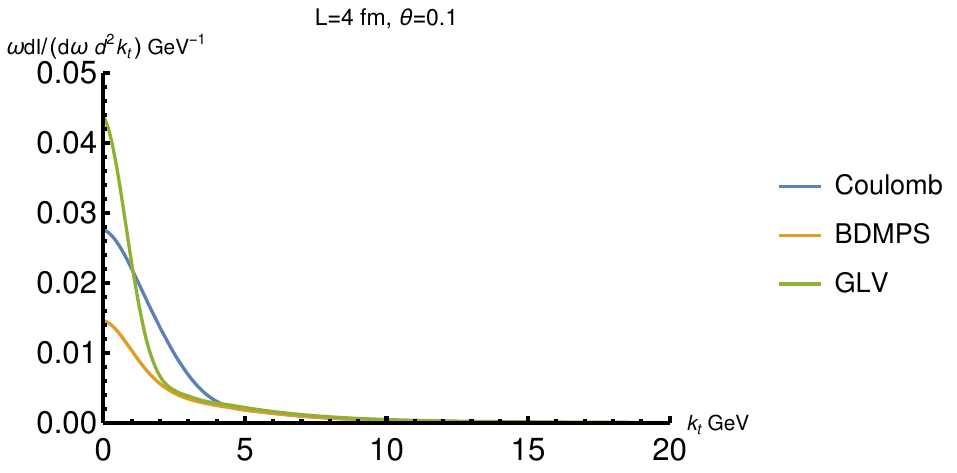}
\includegraphics[scale=0.5]{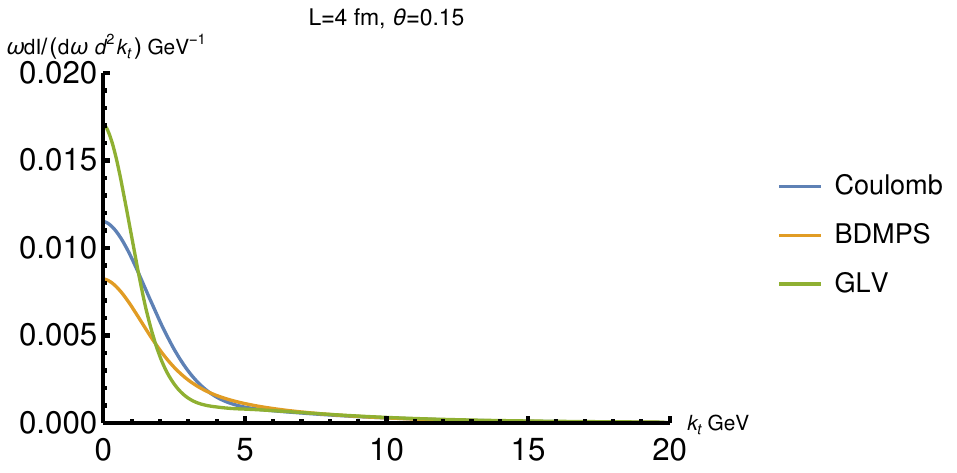}
\vskip 0.8 cm
\center{\includegraphics[scale=0.5]{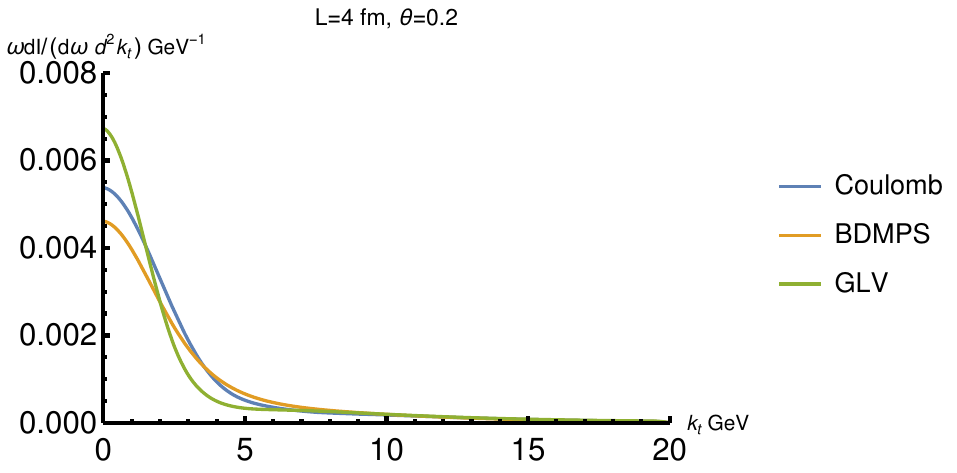}}
\caption{\textsl{The angular distribution of radiated gluons for $\omega= 20 $ GeV } for different $\theta=0,0.05,0.1,0.15,0.2$}
\label{f3}
\end{figurehere}
 We expect  that the sign changing part of the distributions for large $k_t$ is actually an artifact of the soft gluon approximation, that becomes unaplicable for large $k_t$.
 The Coulomb correction is approximately constant at small $k_t$ and starts to decrease  for $k_t$ larger than the BDMPS maximum.
\subsection{Energy loss.}
\par It is also interesting to check how the combined effect of the phase space constraints and Coulomb logarithms  influence the energy loss. We used the soft gluon approximation, so introducing the explicit boundary for the $k_t$ may be beyond the accuracy of our approach \cite{ASW}, but still 
introducing the boundary $k_t\le \omega$ will give a good indication of the effect.
\par It is easy to integrate over $k_t$ in arbitrary finite limits, analytically, since the integrands  in the expressions for angular distributions in the previous two chapters ,
since these expressions are gaussian in $k^2_t$. The remaining integrals are integrals in 
over  $t,t_1,s$  and are taken numerically using Mathematica, in the same way as the integrals for angular distributions.

\vskip 0.8 cm
\begin{figurehere}
\includegraphics[scale=0.5]{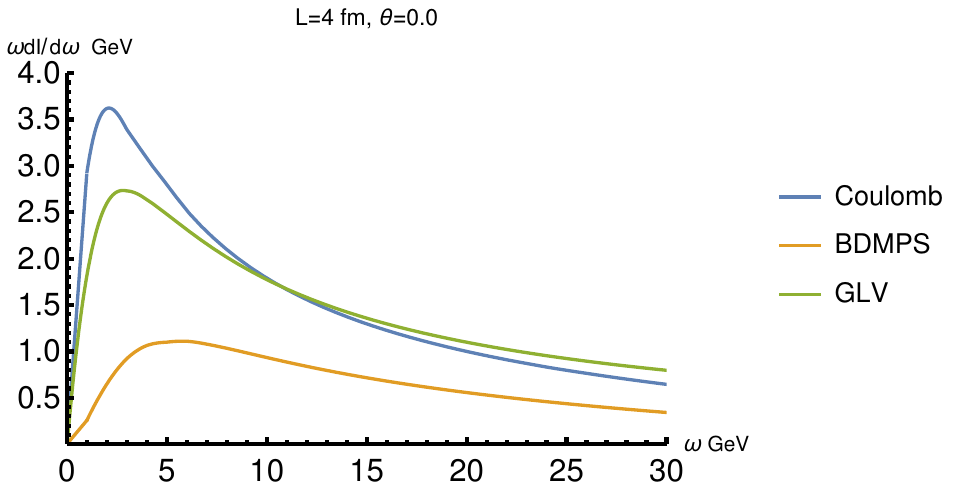}
\includegraphics[scale=0.5]{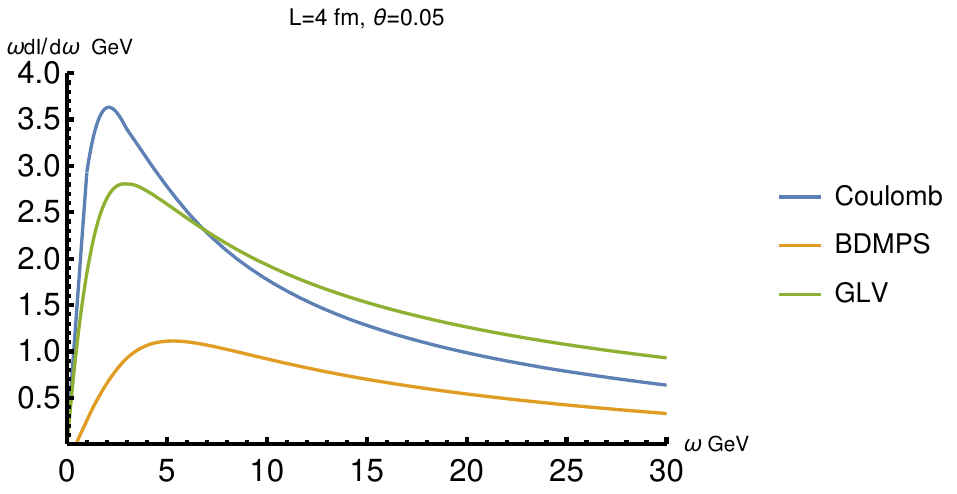}
\vskip 0.8 cm
\includegraphics[scale=0.5]{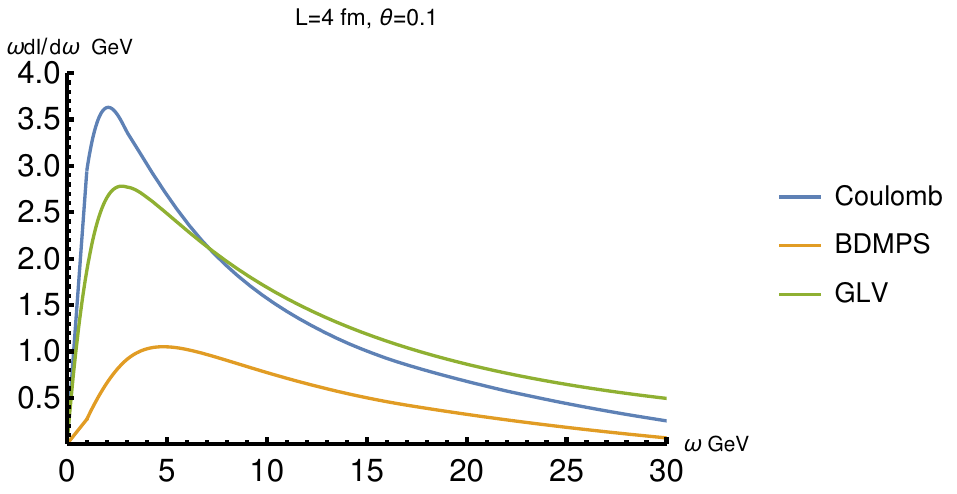}
\includegraphics[scale=0.5]{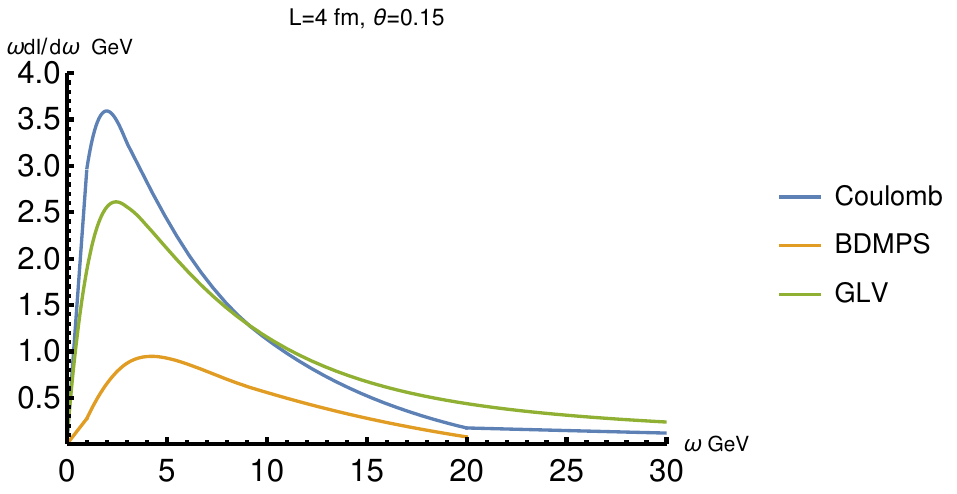}
\vskip 0.8 cm
\centerline{\includegraphics[scale=0.53]{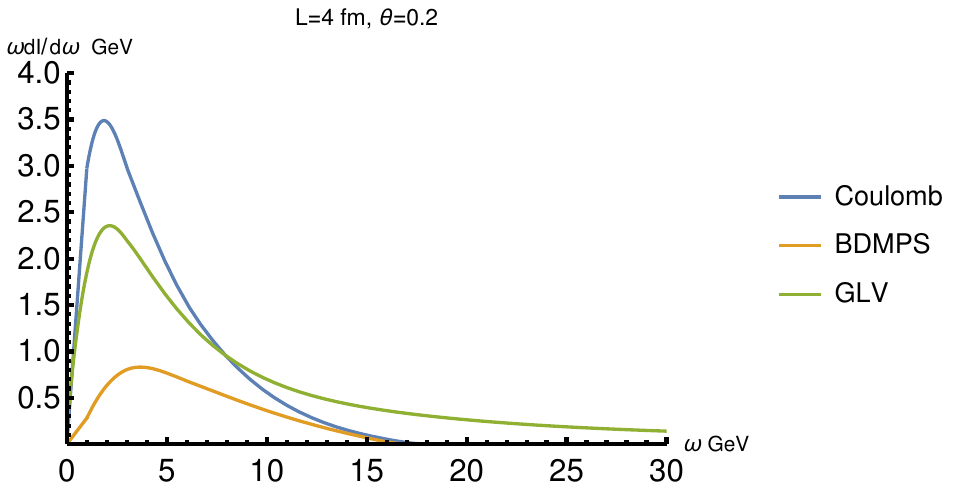}}
\caption{\textsl{The energy loss $\omega dI/d\omega$ with Coulomb gluons  for different $\theta=0,0.05,0.1,0.15,0.2$, The energy loss $\omega dI/d\omega$ was calculated by integrating  the corresponding angular  distribution over $k_t$ in the finite interval of $k_t$ from $k_t=0$  to the kinematical bound
$k_t\le \omega$.  BDMPS means the expression for soft gluon emission in harmonic oscillator approximation, and  Coulomb means the full result including Coulomb  logarithms (Moliere Theory). }}
\label{f4}
\end{figurehere}
We see that compared with the BDMPS spectrum calculated with the same boundary conditions, the  corrections increase the energy loss and is rather close to GLV energy loss.
Note however that this is just the qualitative estimate since, as we remarked above, the angular distributions were calculated in soft gluon approximation, and precise phase space constraints are beyond the accuracy of this approximation \cite{Wiedemann1,ASW}..
\subsection{Quenching weights.}
\par It will be also interesting to estimate the quenching weights  in soft gluon approximation   \cite{ Wiedemann3,ADSW} 
but including the finite integration region in the transverse momentum $k_t\le \omega$ and the  gluons. 
\par As it is known the jet quenching factor describes the energy loss due to the arbitrary number of Poisson distributed gluons.
Indeed, in the previous chapters we calculated the energy loss probability $\omega dI d\omega$ in the first order in $\alpha_s$. 
Then we can calculate the quenching factor
\beq
Q(E)=\exp(-\int^E_0(1-\exp(-\frac{{\it R}}{E}\omega))\frac{dI}{d\omega},\nonumber\\[10pt]
\label{q1}
\eeq
where
 \beq
{\it R}=\frac{d\sigma^{0}}{dp^2_t}
\eeq
is determined from the experimental data, ${\it R}\sim 5$. Here $\sigma^{0}$ is the radiation cross section in the vacuum, outside of the media.
\setlength{\extrarowheight}{0.2cm}
\begin{tablehere}
\begin{center}
\begin{tabular}{cccc}  \hline  \hline 
 & \hspace{0.5cm} $E=25$ GeV \hspace{0.5cm} & \hspace{0.5cm} $E=35$ GeV \hspace{0.5cm} & \hspace{0.5cm} $E=50$ GeV \\  
 BDMPS& - S(E) &- S(E) & -S(E)\\ \hline
Light quark m=0& 0.66 &  0.61&0.53  \\
Heavy quark $m_b=5$ GeV & 0.33     &  0.36      & 0.4\\
BDMPS+Coulomb &  -S(E) &- S(E) & -S(E)\\ \hline
Light quark m=0& 1.8 &  1.6 &1.32  \\
Heavy quark $m_b=5$ GeV & 1.1     &  1.13    & 1.1\\

\hline
\end{tabular}
\caption{\textsl{The estimate for quenching coefficients S(E) for light and heavy quarks, for  $L=4$ fm width. The jet quenching factor $Q(E)=\exp(S(E))$ Here  $\alpha_s=0.3$}. The BDMPS means quenching weight
calculated in the harmonic oscillator approximation with constraint $k_t\le \omega$, $\omega\le E$. BDMPS+Coulomb means quenching weights calculated in the Moliere Theory (BDMPS plus Coulomb  logarithms) with the same limitation $k_t\le \omega$,$\omega\le E$ . }
\label{quenching}
\end{center}
\end{tablehere}
The quenching weights for given energy do not change between $\theta=0$ and $\theta\sim 0.06$. We see that the inclusion of Coulomb  gluons improves the agreement with experimental data,
leading to ratio of quenching weights of massless and heavy quarks with $m_Q=5$ GeV (bottom quark). We have $Q=\exp(-S(E)$ of order 1 at 100 GeV, 0.8 at 50 GeV and 0.65- 0.5  for  35 and 25  GeV, 
There is no difference between massless and charm quarks, at least for jet energies above 25 GeV.

\section{Longitudinal Phase Space Constraints.}
 
\par   In the previous section we studied the heavy quark radiation in the Moliere theory in the soft gluon approximation $k_t\ll \omega$, $\omega \ll E$. It was shown in  \cite{Z0A,Z0B,Z0C,Z1,arnold1,arnold2},
that one can take into account the finite gluon energy.

This means   that for a parton with  the energy $Ez, 0<z<1$ whose propagator we calculate the effective mass in the propagator is substituted from $\omega =Ez$ to $Ez(1-z)$. 
As it was pointed in \cite{Z0A} there is no sense to continue beyond $0\le z \le 1/2$. 
Then the effective potential which was without phase constraints 
\beq
V=\frac{1}{4}\hat q\rho^2\log(1/(\mu^2\rho^2)=\frac{1}{4}\hat q \rho^2\log(Q^2/\mu^2)=\frac{1}{4}\hat q \rho^2\log(1/(Q^2\rho^2))
\label{pot1}
\eeq
becomes 
\beq
V(\rho )=\frac{1}{8}\hat q( \rho^2\log(1/(\rho^2\mu^2)+((1-z)^2\rho^2\log 1/((1-z)^2\rho^2\mu^2)-z^2/9x^2\log(
1/(z^2\rho^2\mu^2)
\eeq
note that in the $z\rightarrow 0$ limit the potential will be   given by Eq. \ref{pot1}.
The expression for angular distribution is now 
 \begin{eqnarray}
 z \frac{dI}{dz d^2k_t }&=&\frac{(1+(1-z)^2)}{2}\frac{C_F\alpha_s}{(2\pi)^2(z(1-z))^2}2Re\int d^2y \int ^\infty_0dt_1\int^{t_1}_0 dt e^{-i\vec k_t\vec y}\nonumber\\[10pt]
&\times& e^{-\int^\infty_{t_1} ds n(s)V(\vec y(s))}\partial_{\vec x}\partial _{\vec y}(K(\vec y,t_1,\vec x,t)-K_0(\vec y,t_1;\vec x,t))\vert_{\vec x=0}.\nonumber\\[10pt]
\label{er1}
 \end{eqnarray}   
 where the propagator for massive quark is now calculated with substitution $\omega\rightarrow Ez(1-z)$  and satisfies 
 \beq
(i\frac{\partial}{\partial t}+\frac{\vec \partial ^2}{2z(1-z)E}+iV(x)+m^2) K(\vec y,t_1;\vec x,t))=i\delta(\vec x -\vec y)\delta (t-t_1)
\eeq
The function n(s)=U(L-s)U(s) is a QGP density profile for the propagating heavy qyar
We split the potential into a sum 
\beq
V=V_0+V_{pert}
\eeq
The potential $V_0$ is now given by 
\beq
V_0(\vec \rho)=\frac{1}{8}\hat q \rho^2(\log(Q^2/\mu^2)+((1-z)^2\log Q^2/((1-z)^2\mu^2)-z^2/9\log(
Q^2/(z^2\mu^2)
\eeq
meaning that the effective coefficient $\hat q$ is given by
\beq
\hat q_{eff}=\hat q\frac{1}{2}(\log(Q^2/\mu^2)+((1-z)^2\log Q^2/((1-z)^2\mu^2)-z^2/9\log(
Q^2/(z^2\mu^2)
\eeq
while the perturbation is now as before given by 
\beq
V_{pert} (\vec\rho)=\frac{\hat q}{8}(1+(1-z)^2-z^2/9)\rho^2\log{1/(Q^2\rho^2)}
\eeq
To obtain the numerical results we just need  to substitute $\omega->Ez(1-z)$ in the results of the previous section,
including the choice of the effective momentum scales \cite{mehtar,mehtar1}. In addition the coefficient in front of ic term 
is given by 
\beq
\frac{\hat q}{8}(1+(1-z)^2-z^2/9)
\eeq

\subsection{Angular distributions}

\par We have depicted the corresponding angular distributions in Fig. \ref{f4} for several values of z and  for energies $E=50$ and 35 GeV.
We chose z=0.1,0.4 for E=50 GeV and  z=0.14,0.5 for E=35 GeV. 
\vskip 1cm
\begin{figurehere}
\includegraphics[scale=0.5]{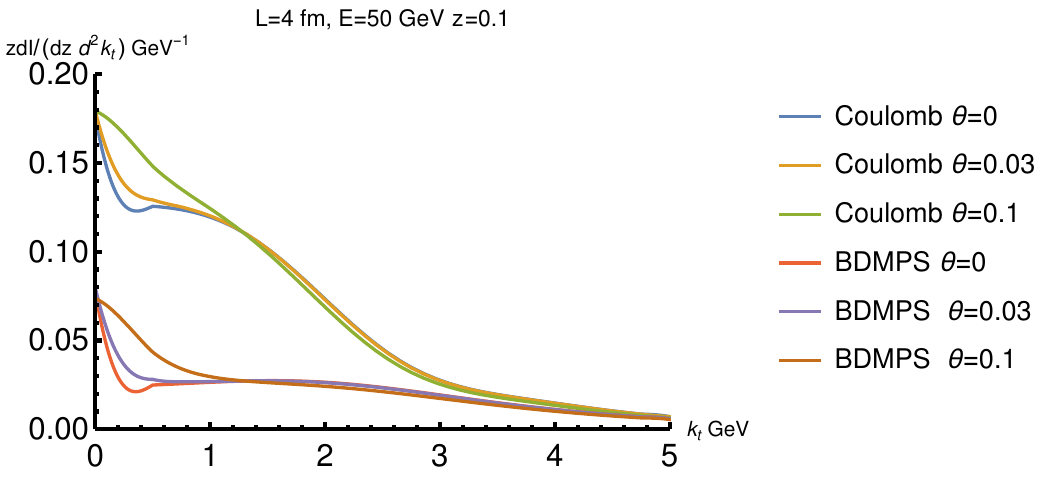}
\includegraphics[scale=0.5]{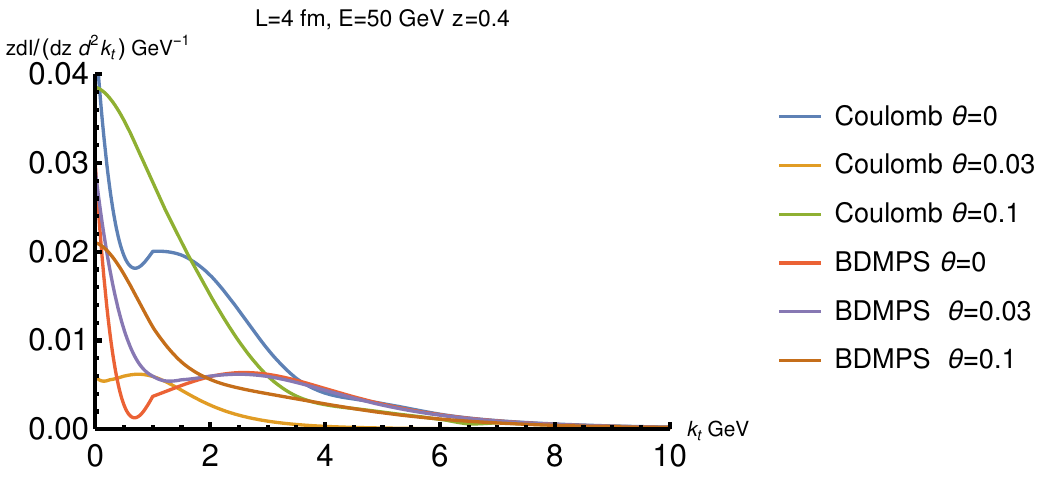}
\caption{The angular distributions  $z\frac{dI}{dzd^2k_t}$   for different $\theta$=0,0.03,0.1 for E=50 GeV (corresponding to massless, charmed and bottom quarks).  As above BDMPS  means the angular gluon distribution 
calculated in the harmonic oscillator approximation but  including finite gluon energy, Coulomb  means the angular distribution  in the Moliere theory (i.e. BDMPS+Coulomb Logarithms) calculated 
taking into account finite gluon energy.} 
\label{f41}
\end{figurehere}
\vskip 1  cm
\begin{figurehere}
\includegraphics[scale=0.5]{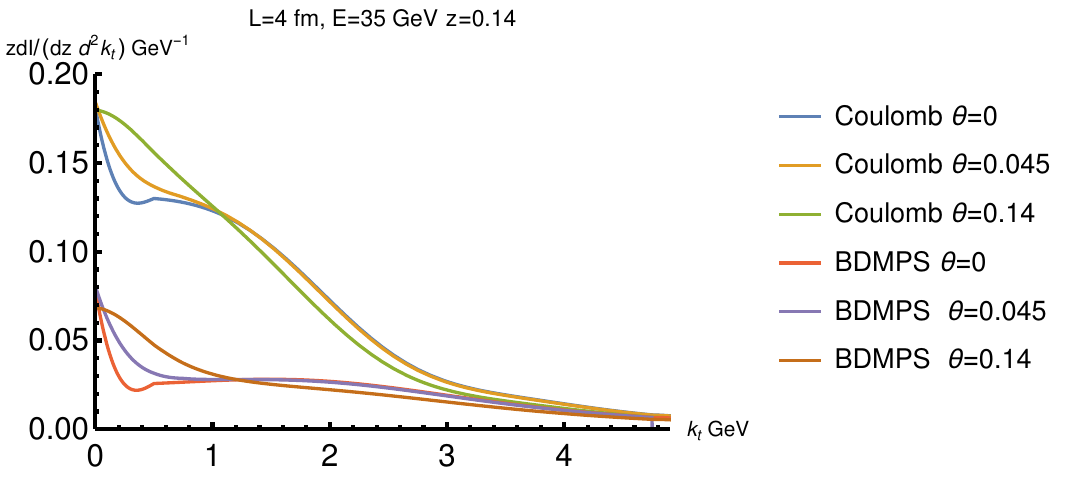}
\includegraphics[scale=0.5]{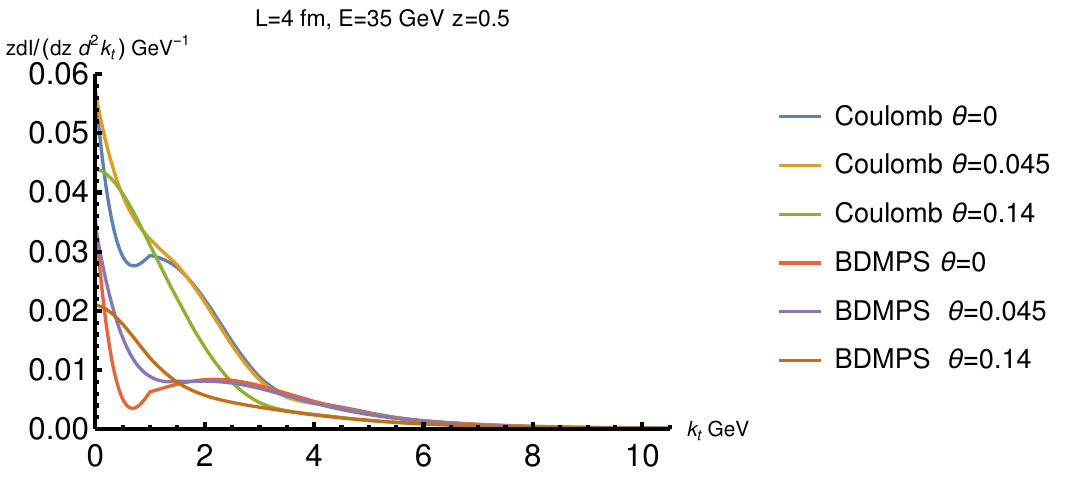}
\caption{The angular distributions  $z\frac{dI}{dzd^2k_t}$   for different $\theta$=0, 0.045,0.14 
(corresponding to massless, charmed and bottom quarks)  for E=35  GeV and z=0.14 (left), z=0.5 (right)} 
\label{f5}
\end{figurehere}
We see that the inclusions of the longitudinal phase space constraints, i.e. the finite gluon energy  significantly improves the behaviour of the angular distributions, but qualitatively the situation is the same
as in the soft gluon approximation: both phase space constraints and  gluons lead to the filling of the dead cone, and the Coulomb gluons give a significant correction to the BDMPS
distributions at small $k_t$.. Note also that the distributions for $\theta=0$ and $\theta=0.05$ practically do not differ, meaning the radiation of the charmed quark is not different from the massless quark.
We chose the values of $\theta$ to have a mass of heavy quark 5 GeV, corresponding to realistic case of the $b$ quark.

\subsection{Energy loss}
The  inclusion of longitudinal phase space constraints also has significantly influences the energy loss. We consider here the spectrum up to $z=1/2$ \cite{Z0A} assuming heavy
quark to be the leading particle. We limit the integration over $k_t$ up to $\omega=Ez(1-z)$. In this way we keep the whole positive value region and cut off the small in magnitude 
tail of the distribution where we expect that  that approximations made in the matrix element calculations may become unreliable.
 We see that the longitudinal phase constraints significantly decrease the influence of the increase in the quark mass. Note that the results between  $\theta=0$ and `$\theta=0.05-0.06$ 
are virtually identical and  thus there is no difference in the energy loss spectrum between light and charm quarks at least for energies at least above 25 GeV.\\
\vskip 1cm
\begin{figurehere}
\centerline{\includegraphics[scale=0.7]{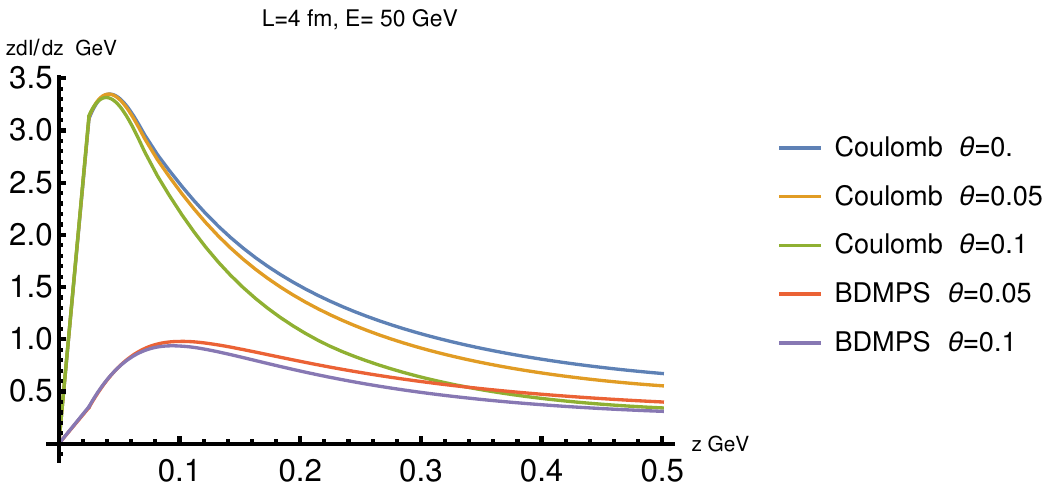}}
\centerline{\includegraphics[scale=0.7]{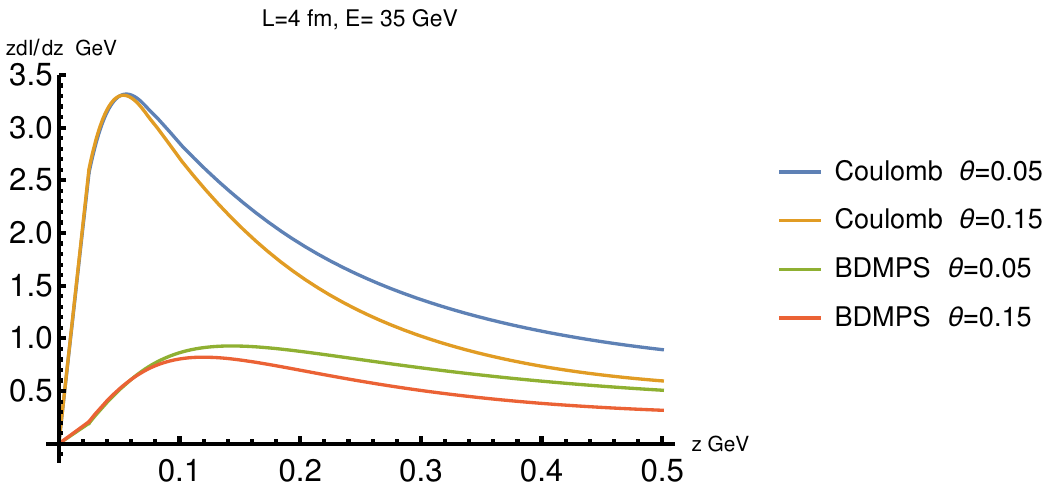}}
\centerline{\includegraphics[scale=0.7]{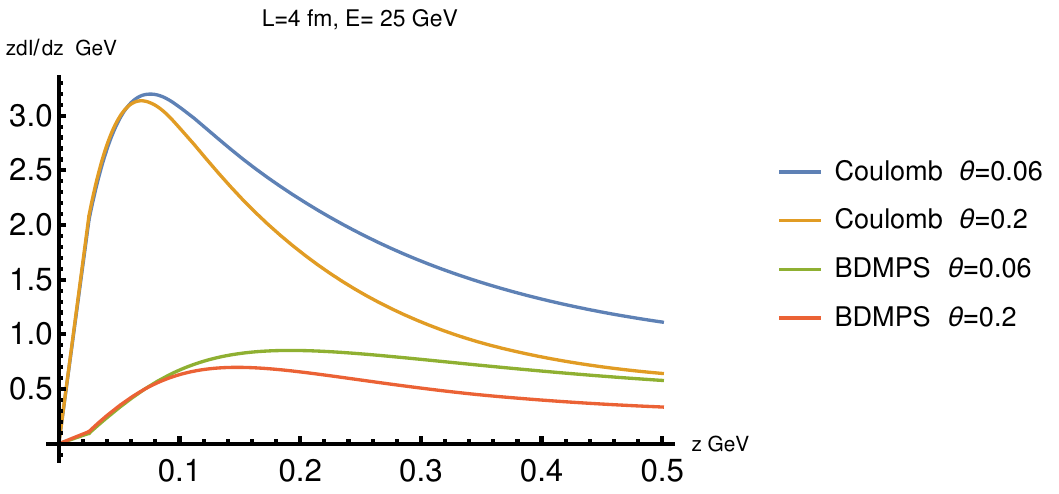}}
\caption{The energy loss  zdI/dz with ic gluons  and BDMPS with phase constraints for different Energies and heavy quark masses $m=1.5$ and 5 GeV.} 
\label{f6}
\end{figurehere}
\vskip 0.8 cm

\subsection{Quenching weights}
We can now calculate the quenching weights and see the significant decrease of the dependence of the quenching weight on the quark mass due to imposition of the longitudinal constraints.

\begin{tablehere}
\begin{center}
\begin{tabular}{cccc}  \hline  \hline 
 & \hspace{0.5cm} $E=25$ GeV \hspace{0.5cm} & \hspace{0.5cm} $E=35$ GeV \hspace{0.5cm} & \hspace{0.5cm} $E=50$ GeV \\  
 BDMPS& - S(E) &- S(E) & -S(E)\\ \hline
Light quark m=0&  0.38&  0.4& 0.4 \\
Heavy quark $m_b=5$ GeV & 0.28     &  0.33      & 0.36\\
 Coulomb&  -S(E) &- S(E) & -S(E)\\ \hline
Light quark m=0& 1.26 & 1.2  &1.01  \\
Heavy quark $m_b=5$ GeV & 1.06     &  1.04    & 1.0\\

\hline
\end{tabular}
\caption{\textsl{The estimate for quenching coefficients S(E) for light and heavy quarks, for  $L=4$ fm widths. The jet quenching factor $Q(E)=\exp(S(E))$. Here  $\alpha_s=0.3, C_F=4/3$}}
\label{quenching1}
\end{center}
\end{tablehere}
Here 
\begin{equation}
S(E)=\int^{1/2}_0 zdI/dz(e^{-nz}-1)/z
\end{equation}
and we assume $\alpha_s=0.3$. Here the quenching weight $Q(E)=\exp(S(E))$.

\section{Conclusions}   
\par We have extended the Moliere theory to angular distributions of the radiation of the heavy quark propagating in the QGP.  We have found for the first time explicit expressions
for Coulomb corrections  to angular gluon distributions in the harmonic oscillator approximation approach (Improved Opacity Expansion ).
We have shown 
that the Coulomb logarithms give a large contribution to harmonic oscillator approximation, with the numerical results indicating that the final answer
is between N=1 GLV and harmonic oscillator approximation. For the case of intermediate widths
considered in the numerical example in this paper (L=4 fm)  the results for Moliere theory are actually rather close to GLV for both transverse distributions and energy loss.
\par Note that for finite quark masses the Coulomb correction is maximal at small $k_t$,
thus enhancing the collinear contribution to the spectrum ("filling the dead cone"). 
This enhancement  ("filling the dead cone") is already present in the Harmonic Oscillator Approximation as it was first noted in \cite{ASW}, and is further enhanced by Coulombic
correction in Moliere theory.
\par We see that the inclusion of transverse phase constraints significantly decreases the dependence of energy loss on the quark mass. This was first noted in \cite{ASW} for the Harmonic Oscillator
Approximation, and persists in the Moliere theory.

Our results indicated that the picture will further improve if  we include  longitudinal DGLAP type phase constraints, that take into account a finite gluon energy. In this case we see that the dead cone effect at energies above 35 GeV is rather small. Further study of the phase space constraints is needed 
to have quantitative agreement with the experimental data . Nevertheless we see that combining phase constraints and Moliere theory we get the results at least qualitatively agreeing with the experimental data,
and making a basis of the construction of the realistic models of the heavy quark energy loss based on Moliere theory \cite{blok2}.
\par After this paper was submitted, a calculation of transverse distributions in the framework of
the improved opacity expansion was presented in \cite{new}. Our results for $\theta =0$ look in agreement with that of \cite{new} for transverse distributions. Some numerical differences may
be related for the use of the single matching scale in the current paper, while the Ref. \cite{new}
uses different matching scales for the exponent expansion and the rest of the spectrum.
The author thanks K. Tywoniuk for the discussion on this subject.
\acknowledgements This work was supported by the Israel Science Foundation  grant  2025311.
\appendix*
\section{}
Here we present explicit expressions for functions $F_2,F_3,F_4$  used in the calculations of the angular distributions.
\beq 
F_1(p,c,Q)=\int^{\infty}_0dx x\log(Qx)J_0(cx)\exp(-px^2)=(1/p)\exp(-c^2/(4p))(\log(cQ/(2p))-0.5Ei(c^2/(4p))
\eeq
\begin{eqnarray}
F_2(p,c,Q)&=&\int^{\infty}_0dx x^3\log(Qx)J_0(cx)\exp(-px^2)\nonumber\\[10pt]
&=&-\frac{1}{p^3}(-p\exp(-c^2/(4p))+p/2\nonumber\\[10pt]
&+&(-c^2/8+p/2)Ei(c^2/(4p))+(c^2/4-p)\log(0.5Qc/p))\exp(-c^2/(4p))\nonumber\\[10pt]
\end{eqnarray}
\begin{eqnarray}
F_3(p,c,Q)&=&\int^{\infty}_0dx x^5\log(Qx)J_0(cx)\exp(-px^2)\nonumber\\[10pt]
&=&-\frac{1}{p^5}(\exp(-c^2/(4p))(3p-c^2/2)p+(c^2/8-3p/2)p+(\exp(-c^2/(4p))(-c^4/32\nonumber\\[10pt]
&+&c^2p/2-p^2)Ei9c^2/(4p)+(c^4/16-c^2p+2p^2)\log(Qc/(2p))\nonumber\\[10pt]
\end{eqnarray}
\begin{eqnarray}
F_4(p,c,Q)&=&\int^{\infty}_0dx x^4\log(Qx)J_1(cx)\exp(-px^2)\nonumber\\[10pt]
&=&\frac{1}{cp^4}((p-c^2/4)p+\exp(-c^2/(4p))(p(3c^2/4-p)\nonumber\\[10pt]
&+&c^2((c^2/16-p/2)Ei(c^2/(4p))+(-c^2/8+p)\log(Qc/(2p)))\nonumber\\[10pt]
\end{eqnarray}

 \end{document}